\documentstyle{aipproc2}

\input epsfc.sty
\def\fdg{\hbox{$.\!\!^\circ$}}

\begin{document}

\title{The SPOrt Project: \\ Cosmological and Astrophysical Goals}
\author{R. Fabbri$^{*}$, S. Cortiglioni$^{\dagger }$, 
S. Cecchini$^{\dagger }$, M.
Orsini$^{\dagger \ddagger }$, E. Carretti${\ddagger}$,\\
G. Boella$^{\P }$, G.
Sironi$^{\P }$, J. Monari$^{\S }$, A. Orfei$^{\S }$, 
R. Tascone$^{\S \S}$, U. Pisani$^{\S *}$, \\
K.W. Ng$^{\Vert }$,
L. Nicastro$^{**}$, L. Popa$^{\dagger \dagger }$, I.A. Strukov$^{\ddagger
\ddagger }$, M.V. Sazhin$^{\P \P }$}
\address{%
$^{*}$Dipartimento di Fisica, Universit\`a di Firenze, Via S. Marta 3,
I-50139 Firenze, Italy \\
$^{\dagger }$I.Te.S.R.E./CNR, Via P. Gobetti 101, I-40129 Bologna, Italy \\
$^{\ddagger }$Dipartimento di Astronomia, Universit\`a di Bologna, Via
Zamboni 33, I-40126 Bologna \\
$^{\P }$Dipartimento di Fisica, Universit\`a di Milano, Via Celoria 16,
I-20133 Milano, Italy \\
$^{\S }$I.R.A./CNR VLBI Radioastronomical Station of Medicina, Via P.
Gobetti 101, I-40129 Bologna, Italy \\
$^{\S \S}$CESPA/CNR c/o Dpt. Elettronica Politecnico di Torino, 
c.so Duca degli Abruzzi 24, 10129 Torino, Italy\\
$^{\S *}$Dpt. Elettronica Politecnico di Torino, 
c.so Duca degli Abruzzi 24, 10129 Torino, Italy\\
$^{\Vert }$Institute of Physics, Academia Sinica, Taipei, Taiwan 11529,
R.O.C. \\
$^{**}$I.F.C.A.I./CNR, Via U. La Malfa 153, I-90146 Palermo, Italy \\
$^{\dagger \dagger }$Institute of Space Sciences, R-76900
Bucharest-Magurele, Romania \\
$^{\ddagger \ddagger }$Space Research Institute (IKI), Profsojuznaja ul.
84/32, Moscow 117810, Russia \\
$^{\P \P }$Schternberg Astronomical Institute, Moscow State University,
Moscow 119899, Russia}
\maketitle

\begin{abstract}
We present the cosmological and astrophysical objectives of the SPOrt
mission, which is scheduled for flying on the International Space Station (ISS)
in the year 2002 with the
purpose of measuring the diffuse sky polarized radiation in the microwave
region. We discuss the problem of disentangling the cosmic background
polarized signal from the Galactic foregrounds.
\end{abstract}

\section*{Introduction}

SPOrt is an experiment selected by the European Space Agency for the
International Space Station and intended to measure the linear
polarization of the diffuse sky radiation at an angular resolution of
$7^{\circ }$ in the 20--90 GHz frequency range. Although it was originally
planned to measure the polarized Galactic background, its present design 
will possibly allow the detection
of the polarization of the Cosmic Background
Radiation (CBR). Differing from the other planned space experiments (MAP 
(1) 
and PLANCK  (2)
), it is specifically designed for a
clean measurement of the Stokes parameters $Q$ and $U$, with no significant
limitation arising from spurious polarization. The instrumental design is
presented in a companion paper in this volume (3)
. Here we
discuss the experiment's expected performance (sensitivity and sky coverage)
and the capability to detect and discriminate between the various
contributions to the sky polarized background. Generally speaking, an
experiment sensitivity derived from instrumental noise alone may not provide
an accurate evaluation of the effective sensitivity for the Stokes
parameters of CBR (or the other backgrounds). This point will be considered
in this paper, deriving in particular a reliable estimate of SPOrt
effective sensitivity to CBR. According to our analysis, taking into account
both the present experimental knowledge and the theoretical models, we can
realistically expect to obtain the following results from SPOrt:

\begin{itemize}
\item  High-accuracy, low-resolution maps of Galactic synchrotron at the
lowest frequencies,

\item  discrimination of low temperature dust in our Galaxy,

\item  and last but not least, the
detection of CBR polarization at 60--90 GHz if
the cosmic medium underwent a secondary ionization and the reheating optical
depth was not too weak, say\textbf{\ }$\tau _{\mathrm{rh}}\gtrsim 0.1.$
\end{itemize}

\section*{SPO\lowercase{rt}'s characteristics}

\begin{figure}
\centerline{%
\epsfysize=7.5cm
\epsfbox[0 385 540 737]{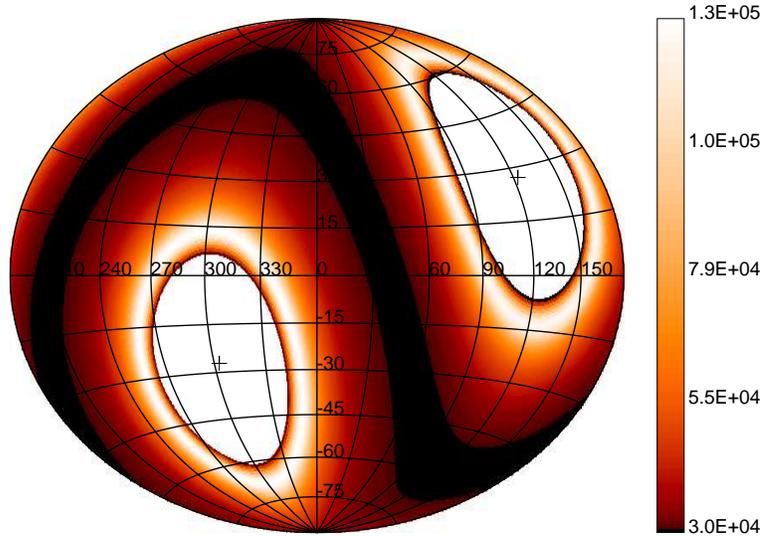} }
\caption{Galactic projection of the SPOrt sky coverage. Exposure times in
 seconds are shown on the right bar.
 Regions around the celestial poles (marked) are uncovered because of pointing
 constraints.}
\label{cover}
\end{figure}

The expected performance for each of the frequency channels (22, 32, 60 and
90 GHz) is summarized in Table \ref{performance}.
Column 2 in the Table gives the mean sensitivity per $7^{\circ }$ pixel,
expressed in terms of the total polarized intensity $P=(Q^2+U^2)^{\frac 12}$.
The integration time per pixel will depend on sky coordinates, ranging
from 29.7 Kilosec (for low values of declination $\delta$) to 128.5 Kilosec
(for the largest values of $\left| \delta \right| $
compatible with the orbit of ISS).
The reported $P_{\mathrm{rms}}$ are computed using actual integration times
and the noise-equivalent temperatures reported by
(3
), the spurious polarization limit being
lower than noise. They also include a 50\% efficiency factor.
%
\begin{table}
\caption[ ]{SPOrt expected performance.} \label{performance}
\begin{tabular}{lddd}
{\boldmath $ \nu$ (GHz)} &
 \multicolumn{1}{c}{%
{\boldmath
$P_{\mathrm{pix}}(\mathrm{ave})$}\protect{\tablenote%
{Computed for 50\% efficiency and 10\% frequency bandwidth.
}} ($\mu $K)}%
& \multicolumn{1}{r}{%
{\boldmath $P_{\mathrm{rms}}$} {\bf (FS)}\protect{\tablenote%
{Full sky (FS) coverage is 81.7\% of 4$\pi $ sr, including 662
pixels.}} ($\mu $K)}
& \multicolumn{1}{r}{{\boldmath $P_{\mathrm{rms}}$}
{\bf (GC)}\protect{\tablenote%
{Galactic cut (GC) excludes a $\pm 20^{\circ }$ belt about the
Galactic plane, retaining 445 pixels.}} ($\mu $K)}
\\
\tableline
22  &  13.4  & 0.52 & 0.64 \\
32  &  13.8  & 0.54 & 0.66 \\
60  &  14.7  & 0.57 & 0.70 \\
90  &  16.8  & 0.65 & 0.81
%
%
\end{tabular}
\end{table}
Column 3 in the Table gives the full-sky sensitivity for the total sky
coverage, which includes 662 pixels and is shown in Figure \ref{cover}. The
last column gives the sensitivity after subtraction of a Galactic plane belt
of $\pm 20^{\circ }$. Clearly the pixel sensitivity $P_{\mathrm{pix}}$ is
intermediate between the expected sensitivities of MAP (1
) and
PLANCK (2
). However the beamwidth is much larger here, so that
the full-sky $P_{\mathrm{rms}}$ cannot be better than 0.5--0.6 $\mu $K for
each channel.

\section*{The expected scenario for Galactic foregrounds}

Generally speaking, both the Galactic and the extragalactic background
should be considered as foregrounds with respect to CBR. However the
extragalactic source contribution should be negligible at our
resolution. Galactic emission includes three contributions related to
different physical processes, namely, synchrotron produced by relativistic
electrons moving in the Galactic magnetic field, free-free (Bremsstrahlung)
arising from interaction between free electrons and ions in a plasma, and
dust emission of thermal origin.

The frequency dependence of synchrotron emission is
usually approximated by a power
law in terms of antenna temperature (but the spectral index depends on
spatial position and also changes with frequency),

\begin{equation}
T_{\mathrm{S}}\propto \nu ^S,\;\;S=-(2.6\div 3.2).  \label{syncS}
\end{equation}
Large sky coverage surveys are only available at low frequencies (4--6
), and extrapolations to the microwave region can be made
through Eq. (\ref{syncS}). From the 1.4 GHz linearly polarized galactic
emission survey of Brouw and Spoelstra
(4) 
 with $S=-3.2$ we get $T_{\mathrm{S}}(30$ GHz$)\simeq 40$ $\mu $%
K. Constraints on the rms fluctuation of the emission are set from CBR
anisotropy measurements. From COBE--DMR (7--8%
) we have a $7^{\circ }$
fluctuation $\delta T_{\mathrm{S}}\lesssim 7$ $\mu $K at 53 GHz. From the Tenerife
experiment Davies and Wilkinson (9%
) are able to derive a $5^{\circ }$--8$^{\circ }$
fluctuation $\delta T_{\mathrm{Galactic}}\lesssim 43$ $\mu $K at
10 GHz in the Northern low emissivity region, so that we can extrapolate $%
\delta T_{\mathrm{S}}(30$ GHz$)\lesssim 2$ $\mu $K; this limit however is not
expected to hold over most of the sky. The intrinsic polarization degree $%
\Pi =P/I$ can be easily predicted for synchrotron, 
\begin{equation}
\Pi _{\mathrm{S}}={(3S+3)}/{(3S+1)},  \label{pisync}
\end{equation}
so that values as high as $\simeq 75\%$ are expected. However misalignment
and smearing effects should reduce the measurable $\Pi _{\mathrm{S}}$, so
that at $7^{\circ }$ a better estimate is probably $\lesssim 30\%$.

Free-free emission seems to be the main source of foreground at $\nu \gtrsim 20$
GHz for CBR \textit{anisotropy} measurements. Its spectral dependence is
accurately described by a power law rather insensitive of spatial position
and frequency

\begin{equation}
T_{\mathrm{FF}}\propto \nu ^F,\;\;F\simeq -2.15.  \label{ffF}
\end{equation}
From COBE--DMR Kogut et al. (7--8
)
it can be obtained $\delta T_{\mathrm{FF}}(53$
GHz$)=7\pm 2$ $\mu $K at $|b|>20^{\circ }$ and some correlation with dust,
i.e., with DIRBE. From Tenerife the extrapolated limit $\delta T_{\mathrm{FF}%
}(30$ GHz$)\lesssim 4$ $\mu $K is derived (9
), which again should
not be considered as typical for the whole sky. Oliveira et al. (10) 
 find DIRBE--Saskatoon cross-correlations and derive $\delta T_{%
\mathrm{FF}}(40$ GHz$)=17\pm 10$ $\mu $K at 1$^{\circ }$. Free-free emission
can be polarized via Thomson scattering within optically thick plasma
regions (11)%
. At microwave wavelengths HII regions are to be
considered optically thin; thus estimating an $\Pi _{\mathrm{FF}}\lesssim 5\%$ is
probably a conservative upper limit.

Dust emission is modelled with power law with index 1.5$\div 2,$ or with a
greybody, or following Wright et al. (12)
, with a mixture of
two greybodies with emissivities $\simeq 2$,

\begin{equation}
T_{\mathrm{D}}\propto \nu ^{D-2}\left[ B_\nu (20.4\text{ K})+6.7\cdot B_\nu
(4.77\text{ K})\right] ,\;\;D\simeq 2.  \label{dustD}
\end{equation}
This emission is obviously better known at high frequencies. IRAS and DIRBE
data at $\nu \geq 10^3$ GHz are usually utilized for templates
or cross-correlations (7,10) 
rather than extrapolations to the microwave region. From COBE--DMR (8) 
 we have $\delta T_{\mathrm{D}}(53$ GHz$)=2.7\pm 1.3$ $\mu $K,
which implies $\delta T_{\mathrm{D}}(100$ GHz$)\sim 10$ $\mu $K. Dust
emission can be polarized (provided dust grains are aligned by the Galactic
magnetic field), probably to a level of $\sim 10\%$ (13)
. Sethi et
al. (14) 
 have recently provided a detailed modelling for the
angular spectrum of polarized dust emission. Using the Leiden--Dwingloo HI
maps and a relation between dust optical depth and HI column density, they
compute the microwave emission over the sky, and from a model of spheroidal
silicate-graphite grains derive an intrinsic $\Pi _{\mathrm{D}}$ of 30\%.
Finally modelling the Galactic magnetic field they compute the polarization
reduction factor. From their results we derive the estimate $P_{%
\mathrm{D}}\simeq 0.05$ $\mu $K at $\nu \simeq 100$ GHz on scales of
$7^{\circ }$. This is considerably lower than the estimate from COBE--DMR with
$\Pi _{\mathrm{D}}=10\%$. The discrepancy is indicative of the existing
uncertainties on dust polarized emission at microwave frequencies.
\begin{figure}[t]
\centerline{%
\epsfysize=5.5cm
\epsfbox{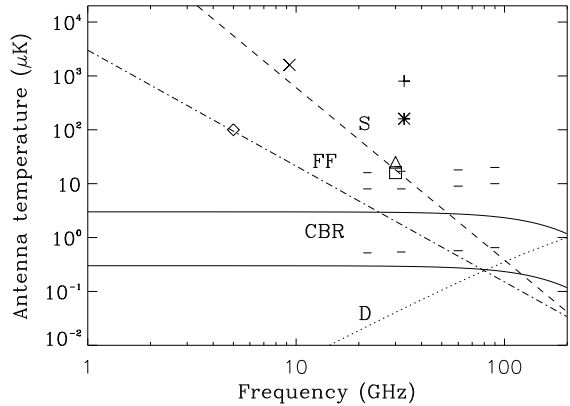} }
\caption{Expected Galactic foregrounds in the microwave region. Synchrotron
(S) and free-free (FF) polarized emissions are normalized to
18 $\mu$K and 2 $\mu$K, respectively, at 30 GHz, and dust (D) to 1 $\mu$K
at 100 GHz.
Also reported are the CBR signal normalized to 1\% and 10\% of anisotropy,
the experimental upper limits quoted in Table 2 and the SPOrt sensitivities.
For each frequency channel we give the maximum and minimum $P_{\mathrm{pix}}$,
and $P_{\mathrm{rms}}$.}
\label{foreg}
\end{figure}

In Fig. \ref{foreg} we provide estimates for the Galactic foregrounds
compared with SPOrt sensitivity limits. Clearly synchrotron emission can
easily be measured at the lowest frequencies. Free-free emission, on the other
hand, is not expected to dominate at any frequency in the 20--90 GHz range.

\section*{CBR polarization}

\subsection*{Present experimental status}

In the last two decades published results on CBR polarization show an
improvement in sensitivity of roughly one order of magnitude, which was not
sufficient to give a positive detection. Table \ref{experim} reports the
available upper limits on $\Pi _{\mathrm{rms}}$, the rms polarization degree
at the corresponding scales $\alpha $ for the sky coverages given in the
4-th column. The most stringent limit is 16 $\mu $K on a scale of 1\fdg4,
for a fairly wide window about the North Celestial Pole (21)
. It should be noted that the upper limits on the full sky $\Pi _{%
\mathrm{rms}}$ would be even weaker than those reported in Table 2.

\begin{table}[b]
\caption[ ]{Upper limits on CBR polarization degree.} \label{experim}
\begin{tabular}[t]{lrrrr}
{\bf Ref.} & {\boldmath $ \nu$ }  (GHz) &{\boldmath $\alpha $} & \multicolumn{1}{r}%
{{\bf Sky coverage}\tablenote%
{GC = Galactic Center, NCP = North Celestial Pole.}} & {\boldmath $\Pi _{%
\mathrm{rms}}$} ($\leq )$ \\  \tableline
(15)
& 4.0 & 15$^{\circ }$ & scattered & 0.1 \\
(16)
& 100--600 & 1\fdg5--40$^{\circ }$ & GC & (10--1)$\times
10^{-4}$ \\
(17)
& 9.3 & 15$^{\circ }$ & $\delta =+40^{\circ }$ & 6$\times
10^{-4}$ \\
(18)
& 33 & 15$^{\circ }$ & $\delta \in (-37^{\circ },+63^{\circ })$
& 6$\times 10^{-5}$ \\
(19)
& 5.0 & $18''$--$160''$ & $\delta =+80^{\circ }$ & (1.4--0.4)$%
\times 10^{-4}$ \\
(20)
& 26--36 & 1\fdg2  & NCP & 1$\times 10^{-5}$ \\
(21)
& 26--36 & 1\fdg4  & NCP & 0.6$\times 10^{-5}$ \\ 
\end{tabular}
\end{table}

As we shall discuss in the next Section, the goal must be a sensitivity
better than a few $\mu $K at $\alpha <1^{\circ }$ and better than a few
tenths of $\mu $K at $\alpha >1^{\circ }$. Several experiments are in
preparation or in the course of execution (22,23,11) 
 at
angular scales ranging from a few arcminutes to 14$^{\circ }$. The best
prospects for positive detection, however, are linked to space experiments (%
1--3).

\subsection*{Theoretical predictions}

The main sources of CBR polarization are the metric perturbations of
spacetime, which include scalar (density), vector (velocity) and tensor
(gravitational) waves; such perturbations directly generate anisotropy, and
polarization is thereby excited by the anisotropy quadru\-pole by the
intervention of Thomson scattering in the cosmic medium. Solutions of the
transfer equation for polarized radiation in the cosmological environment
have been studied for a number of models; see refs. (24, 25),  
and also (26) 
for a short review.
Theoretical predictions from perturbation models are often expressed in
terms of angular power spectra, which are connected to the harmonic
expansions

\begin{eqnarray*}
\Delta T &=&T_0\sum_{l,m}a_{T,lm}Y_{lm}(\vartheta ,\varphi ), \\
Q\pm iU &=&T_0\sum_{l,m}\,_{\pm 2}a_{P,lm}\;_{\pm 2}Y_{lm}(\vartheta
,\varphi ),
\end{eqnarray*}
involving the scalar and spin-weighted spherical harmonics, respectively
denoted by $Y_{\ell m}$ and $_{\pm 2}Y_{\ell m}$ (27)
. The
polarization spectrum is also altered substantially by gravitational lensing
(28) 
 and inhomogeneous reheating (29) 
 for $l > 10^3$; on SPOrt angular scale, however, only spacetime
perturbations are
effective. Since the even (or electric) and odd (or magnetic) parities are
separated through the linear combinations $
a_{E,lm}=-\left( \,_2a_{P,lm}+\,_{-2}a_{P,lm}\right) /2 $ and 
$a_{B,lm}=i\left(
\,_2a_{P,lm}-\,_{-2}a_{P,lm}\right) /2$, 
four angular power spectra are usually provided, 
\begin{equation}
C_{Xl}=\left\langle a_{X,lm}^{*}a_{X,lm}\right\rangle
,\;\;C_{Cl}=\left\langle a_{T,lm}^{*}a_{E,lm}\right\rangle ,  \label{power}
\end{equation}
with $X=T$, $E$ and $B$. (Cross-correlations other than $C\equiv T \times E$ vanish
identically.) The most obvious source of polarization, i.e. the density
perturbation which originated the observed cosmic structure, only produce
E-parity multipoles. E- and B-parity mixed fields are produced by all of the
other sources, and unfortunately, by Galactic foregrounds (e.g., (14)
).

The rms anisotropy and polarization at a given angular scale $\alpha $ can
be estimated by $\Delta T_{\mathrm{rms}}(\alpha )\sim T_0\sqrt{T_{l^{*}}}$
and $P_{\mathrm{rms}}(\alpha )\sim T_0\sqrt{P_{l^{*}}}$, where 
\begin{equation}
T_l={l(l+1)C_{Tl}}/{(2\pi )},\;\;P_l={l(l+1)C_{Pl}}/{(2\pi )},\;
\label{p-estimate}
\end{equation}
with $C_{Pl}=C_{El}+C_{Bl}$ and $l^{*}\sim 180^{\circ }/\alpha $, so that
the quantities defined by Eq. (\ref{p-estimate}) are usually plotted for
power spectra. Examples are provided in Fig. \ref%
{angspct}. Full expressions of the polarization cross-correlation functions
(including the autocorrelation $P_{\mathrm{rms}}(\alpha )$ as a simple case)
in terms of the angular power spectra are given by Kamionkowsky et al. (30) 
 and Ng and Liu (31).

\begin{figure}
\centerline{%
\epsfysize=7.7cm
\epsfbox{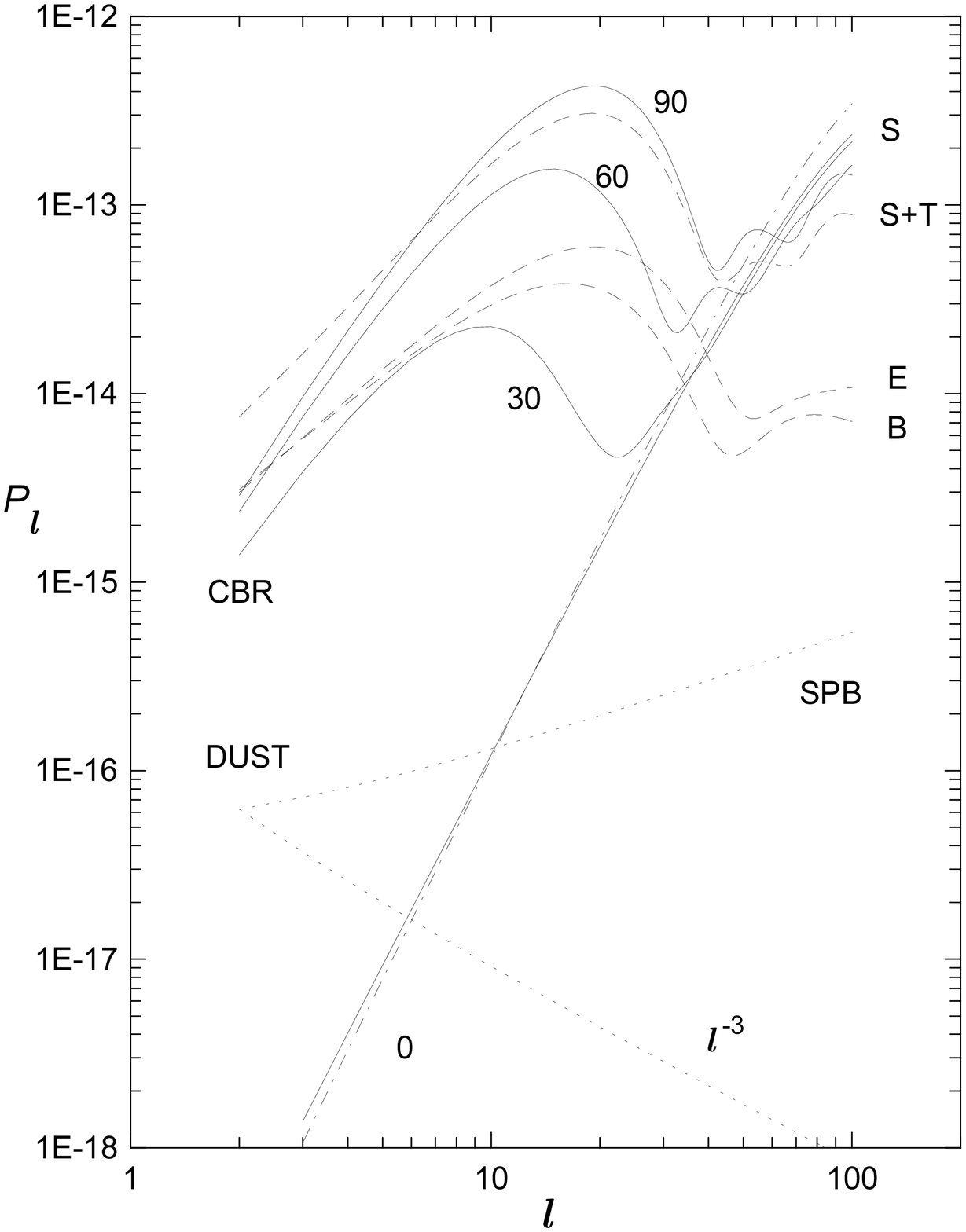} }
\caption{Polarization angular spectra from models of cosmic structure
and Galactic dust emission.
Full lines describe the CBR polarization in SCDM models with a baryon density
$\Omega_{\mathrm{b}}=0.03$ and a scale-invariant spectrum of scalar (S) waves,
and are labelled by the corresponding values of $z_{\mathrm{rh}}$. Dashed lines
refer to  E- and B-parity contributions from tensor waves (labelled by E and
B respectively), and to the total scalar-plus-tensor (S+T) polarization,
normalized so as to get the same anisotropy quadrupole as in the scalar case.
The dash-dotted
curve considers a tilted $n=1.2$ spectrum of scalar waves in the no-reheating
model. Finally, the dotted curves describe the dust emission
model of ref. (14)
(SPB) and the $l^{-3}$ law referring to
foreground anisotropies. }
\label{angspct}
\end{figure}

Standard recombination models, where the CBR photons were last scattered at
a redshift $%
z_{\mathrm rh}\approx 1000$, predict low
levels of polarization at angular scales $\alpha %
\gtrsim 1^{\circ }$. Calculations performed in the SCDM model
(the CDM model with
the total density parameter $\Omega _0=1$ and the reduced Hubble constant $%
h=1$) show that on a scale of $7^{\circ }$ signals of order 0.05 $\mu $K are
expected. Because of a significant increase in the $P_l$ spectrum
beyond the first temperature Doppler peak ($l\approx 200)$, at scales below
$1^{\circ }$ the polarization-to-anisotropy ratio is of order 10\%, so that
we expect polarized signals as high as $\sim 5$ $\mu $K. The prospects for
detection at larger scales  are
more favourable if the cosmic medium underwent a secondary
ionization (reheating). In secondary ionization
models polarization at angular scales $%
\alpha \gtrsim 1^{\circ }$ is produced at a new last-scattering surface placed at
a redshift $z_{\mathrm{ls}}\approx 100\Omega _0^{1/3}\left( 0.025/X\right)
^{2/3}$, with $X=x_{\mathrm{e}}\Omega _bh\;$depending on the ionization
degree $x_{\mathrm{e}}$, the baryon density parameter $\Omega _b$ and the
reduced Hubble constant $h$, or at the reheating onset $z_{\mathrm{rh}}$ if $%
z_{\mathrm{rh}}<$ $z_{\mathrm{ls}}$. Thus a characteristic angular scale
(corresponding to the horizon size at $z_{\mathrm{rh}}$ or $z_{\mathrm{ls}}) 
$ is introduced, which turns out to be in the degree range. The polarization
spectrum is now peaked at such a scale, and suppressed at smaller scales.

The relevant parameters for secondary ionization models are the reheating
redshift $z_{\mathrm{rh}}$,  $X$, $\Omega _0$, and the parameters
of perturbation spectrum (the amplitude, the primordial spectral index $n,$
and the shape factor $\Omega _0h$ for CDM models)$.$ The most important
combination of parameters is given by the approximate expression of the
reheating optical depth for Thomson scattering
\begin{equation}
\tau _{\mathrm{rh}}\simeq 3.8\times 10^{-2}X\Omega _0^{-1/2}z_{\mathrm{rh}%
}^{3/2}.  \label{taurh}
\end{equation}
The amplitude of the main peak in the angular spectrum $P_l$ is roughly
proportional to $\tau _{\mathrm{rh}}^2$ for $\tau _{\mathrm{rh}}\lesssim 1$,
although it also depends on other parameters such as the perturbation
spectral index; its position scales as $l\propto \tau _{\mathrm{rh}%
}^{1/3}X^{-1/3}$. These properties are clear in Figure \ref{angspct} which
gives angular spectra for a few CDM models. For $z_{\mathrm{rh}}=90$
(corresponding to $\tau _{\mathrm{rh}}\simeq 0.5)$ we find $T_0^2P_{l,\max
}\approx (1.8$ $\mu $K$)^2$ at $l_{max}\simeq 20$. Slightly higher levels
of polarization may be found with more favourable spectral shapes, in
particular increasing the primordial index $n$. (This effect is more
evident if models are normalized according to the power-spectrum
quadrupole $Q_{\mathrm rms-PS}$.)  However
as we are going to
discuss below, CBR signals larger than 1 $\mu $K,
although they  cannot be excluded, are not very
likely to occur.

\subsubsection*{Experimental constraints and theoretical expectations for the
reheating strength}

While the Gunn--Peterson test on high-redshift objects implies $z_{\mathrm{rh}%
}\gtrsim 5$, it is more difficult to set stringent upper limits on the strength
of reheating. One might expect that significant constraints on reheating
should come from upper limits on the spectral distortions of CBR. However
the COBE--FIRAS limit on the (generalized) Comptonization
parameter, $y_{\mathrm{C}}<1.5\times 10^{-5}$, cannot exclude a large $z_{%
\mathrm{rh}}$; even no-recombination scenarios are possible if reionization
is non-thermal (32). 
 More significant constraints come from
anisotropy data. The very existence of first Doppler peak implies
$\tau_{\mathrm{rh}} \lesssim 1$%
, and a more significant result comes from the bulk of data on the harmonic
spectrum. Fitting intermediate-scale anisotropy data with the standard
SCDM model, de Bernardis et al. (33) 
 give a best value $%
z_{\mathrm{rh}}\approx 20$, and an upper limit which depends on $\Omega
_bh^2 $ and can be as large as $\sim 100$ if $\Omega _bh^2=0.0075$.

This result can be used to set constraints on $\tau _{\mathrm{rh}}$ when
combined with data on baryon density. Constraints on $\Omega _b$ are
obviously given by the nucleosynthesis requirements. Olive (34) 
 derives $\Omega _bh^2=0.006_{-0.001}^{+0.009}$ from $^4$He and $^7$Li data,
but two mutually inconsistent results, $0.005<\Omega _bh^2<0.014$ or $%
0.017<\Omega _bh^2<0.022$ from Deuterium abundance, according as to whether
one accepts a high (35) 
 or low (36) 
 D/H ratio.
Fukugita et al. (37) 
 compute a cosmic baryon budget from estimates
of known contributions. From their results we derive the best estimate $%
\Omega _bh=0.013+0.001h^{1/2}$, and an upper limit $\Omega
_bh=0.025+0.003h^{1/2}$. All of these constraints appear to be mutually
consistent, with the exception of the Deuterium limit
in the low-D/H case. From the cosmic baryon
budget and the results of de Bernardis et al. (33), for $\Omega _0=1$ we can
derive the upper limit $\tau _{\mathrm{rh}}\lesssim 0.7$, and a best estimate $%
\tau _{\mathrm{rh}}\sim 0.05$.

A low reheating optical depth is supported by explicit modelling of the
ionizing mechanisms. Although a very small fraction of collapsing baryons
would be enough to reionize the rest of the universe at early times, say at $%
z\sim 30$ or higher, it seems difficult to trigger
the formation of stars or QSO black holes which should
allow an early release of UV ionizing radiation.
There was an early condensation of baryonic
objects with $M\sim 10^5M_{\odot }$
according to the CDM scenario, but an insufficient virialization
temperature prevented further fragmentation or collapse of such condensed
objects through H$_2$ cooling. Thus one must wait until objects with $M\gtrsim %
10^8M_{\odot }$ condense. Two different scenarios for the subsequent
evolution of such objects are investigated by Haiman and Loeb (38,39)
. The first considers bursts of star formation with a universal mass
function. The efficiency of star formation is calibrated by requiring\ that
the resulting metallicity at $z\approx 3$ is roughly as observed, $Z\sim
10^{-2}Z_{\odot }$. The other scenario considers the production of
low-luminosity, short-lifetime QSO's with a universal light curve. The
background cosmology here (39) 
is not standard CDM, but the so-called ``concordance''
model, where the cosmological constant provides about 2/3 of the critical
mass, but this makes no important difference. The basic requirement is now
to match the observed luminosity function\ at $z\lesssim 5$. In both cases the
growing number of expanding HII bubbles fills up the universe before\textbf{%
\ }$z\sim 10$, and the resulting optical depth is $\tau _{\mathrm{rh}}\simeq
0.05\div 0.1$ for stars and $\simeq 0.05$ for QSO's.

We can thereby conclude that the polarized CBR signal at $7^{\circ}$ is most
likely to be found in the sub-$\mu$K range.

\section*{The separation of foregrounds}

The separation of the contributions to polarized signals must be based on
the different spectral and spatial behaviours of foregrounds and CBR.
Multifrequency observations thereby play a fundamental role. We performed a
preliminary analysis taking advantage of Dodelson's analytical formalism (%
40)
. This formalism allows us to estimate the experiment effective
sensitivity for any signal component after subtracting other
contributions
on a single-pixel basis (or alternatively, on a single-mode basis after a
spherical harmonic expansion). Let us
consider an experiment measuring the total signal at
frequencies $\nu _1$, $\nu _2$, $\ldots \nu _m$ and intended to extract the
CBR and $n_{\mathrm f}$ foregrounds.
Introducing the $m$-dimensional vectors $\mathbf{S}$%
, $\mathbf{N}$ and ${\mathbf{F}}^{i}$ for the detected signal, instrumental noise
and foreground shapes respectively, we have  $
{\mathbf{S}}=\sum_{i=0}^{n_{\mathrm f}}
s^i{\mathbf{F}}^i+{\mathbf{N}}$, 
with $i=0$ denoting CBR, and the amplitudes $s^i$ to be determined with the experiment.
The problem is at which accuracy we can determine $s^i$ when the noise is
completely described by the auto- and cross-correlation coefficients $%
C_{jk}=\left\langle N_{\nu _j}N_{\nu _k}\right\rangle$. Focusing our
attention on CBR and neglecting cosmic variance, the effective
variance for one pixel or mode amplitude  is given by
\begin{equation}
\sigma _{\mathrm{cbr}}^2=(FDF)^2\left[ \sigma ^{(0)}\right] ^2+\sigma _{%
\mathrm{shape}}^2,  \label{sigmadod}
\end{equation}
where the foreground degradation factor $FDF$ describes the experiment
sensitivity after removal of foregrounds with perfectly known spectral
shapes, and $\sigma _{\mathrm{shape}}$ takes into account uncertainties in
the spectral shapes. The computation of these parameters requires a matrix 
$K^{ij}$ depending on  ${\mathbf{F}}^i$ and noise cross-correlations (40)
. 
If all channels have equal and uncorrelated noise $\sigma _{1ch}$, then$%
\;\sigma ^{(0)}=\sigma _{\mathrm 1ch}/\sqrt{{\mathrm Nch}}$ and  $K^{ij}  
=\sum_{k,l}F_{\nu _k}^iF_{\nu _l}^j$. The final results provided by
Dodelson are $%
FDF=\sqrt{\left( K^{-1}\right) ^{00}}   $
and 
\begin{equation}
\sigma _{\mathrm{shape}}^2=\left\{ \left( \sum_{i=1}^{n_{\mathrm f}}{\mathbf{S}}^i\right)
\cdot \left[ \sum_{j=0}^{n_{\mathrm f}} \left( K^{-1}\right) ^{0j}{\mathbf{F}}^j\right]
\right\} ^2,  \label{sigmash}
\end{equation}
where ${\mathbf{S}}^i$ are the true foreground contributions to $\mathbf{S}$,
whose spectral shapes may differ from the assumed ${\mathbf{F}}^i$. Systematic
contributions to $\sigma _{\mathrm{shape}}$ may come from neglecting the
contribution of some foreground component in the ${\mathbf{F}}^j$ summation in
Eq. (\ref{sigmash}). The best effective sensitivity for CBR is reached by a
proper balance of two contrasting effects, since increasing the number of
fitted foregrounds makes $\sigma _{\mathrm{shape}}^2$ smaller but $FDF$
larger.

\begin{figure}
\centerline{%
\epsfysize=5.5cm
\epsfbox{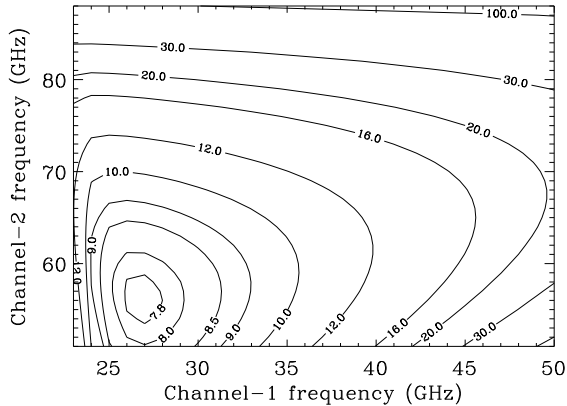} }
\caption{$FDF$ isocontours in the ($\nu _1, \nu _2 $) plane for fixed
 extremal frequencies
 20 and 90 GHz. The minimum $FDF$ is close to SPOrt intermediate frequencies.}
\label{fdfcnt}
\end{figure}

The parameter $FDF$ only depends on spectral shapes and frequency channels.
The $FDF$ level contours shown in Fig. \ref{fdfcnt} show that for the spectra given by Eqs.
(1)--(4) the channel configuration of SPOrt is nearly optimal for the
assigned range (20--90 GHz).
Table \ref{fdfsig} reports the values of the above parameters for different
treatments of SPOrt output. Column 1 in the Table gives the number of
channels used for the foreground fits and the frequency range. When the
number of channels used is 3, the 90 GHz channel is not used; for the
configuration labelled by 4*(22--60), which is not pertinent to the SPOrt
present design, the computation assumes two 60-GHz channels. The second and
third column specify whether free-free and dust emission are fitted. The
results labelled by (1) (5-th and 6-th column) assume polarized intensities
of 18 $\mu$K and 2 $\mu$K for synchrotron and free-free, respectively,
at 30 GHz,
and the assumed spectral slopes are $S=-3.2$ and $F=-2.15$; dust polarized
emission is normalized to 1 $\mu $K at 200 GHz. For the results
labelled by (2)
we normalize synchrotron to 12 $\mu$K and  dust to 0.15 $\mu $K at
the same frequencies as above. The quantities $\sigma _{\mathrm cbr}^{(1)}$
and $\sigma _{\mathrm cbr}^{(2)}$, although they refer
to the total sky coverage,
are computed treating
$\sigma _{\mathrm{shape}}$
as a systematic error.
Clearly the best results
are
found when free-free and dust emission are not fitted (so that $\sigma _{%
\mathrm{shape}}^2\neq 0$). Using four channels we have $FDF=1.38$ and $%
\sigma _{\mathrm{shape}}^{(1)}=0.61$ $\mu $K, and
using SPOrt total-coverage instrumental
sensitivity  from the above
numbers we get $\sigma _{\mathrm{cbr}}^{(1)}$ = 0.73 $\mu $K.
For  $\sigma _{\mathrm{cbr}}^{(2)}$ we can find some results even lower
than 0.5 $\mu$K.

\begin{table}
\caption{The $FDF$ and $\sigma _{\mathrm shape}$ for various
foreground treatments.} \label{fdfsig}
\begin{tabular}{lccddddd}
{\bf Configuration} & {\bf FF} & {\bf Dust} & \multicolumn{1}{c}{\boldmath $FDF$}
&\multicolumn{1}{c}{\boldmath $\sigma _{\mathrm{shape}}^{(1)}$} &\multicolumn{1}{c}
{\boldmath $%
\sigma _{\mathrm{cbr}}^{(1)}$}
& \multicolumn{1}{c}{\boldmath $\sigma _{\mathrm{shape}}^{(2)}$}
& \multicolumn{1}{c}{\boldmath $\sigma _{%
\mathrm{cbr}}^{(2)}$} \\    \tableline
4 (22--90) & Yes & Yes & 8.80 & 0.00 & 2.52 & 0.00 & 2.52 \\
4 (22--90) & No  & Yes & 2.76 & 0.77 & 1.10 & 0.64 & 1.02 \\
4 (22--90) & No  & No  & 1.38 & 0.61 & 0.73 & 0.25 & 0.47 \\
3 (22--60) & No  & No  & 1.54 & 0.64 & 0.80 & 0.17 & 0.51 \\
3 (22--60) & Yes & No  & 3.60 & 0.24 & 1.16 & 0.24 & 1.16 \\
4* (22--60) & No & No  & 1.37 & 0.59 & 0.70 & 0.18 & 0.42 \\
4* (22--60) & Yes & No  & 3.21 & 0.24 & 0.92 & 0.24 & 0.92
\end{tabular}
\end{table}

Slightly higher numbers are found using three channels for the synchrotron
fit. The conclusion is that if dust emission is low, then the fourth channel
adds little to the determination of CBR polarization in a single-pixel
analysis; but if dust is important, it should probably not be included in a
fitting together with synchrotron. A good strategy for data analysis may be
the following, to use three channels up to 60 GHz for synchrotron and CBR
polarized signals, and the 60 and 90 GHz channels for dust and CBR. With
this approach it may be possible to investigate the roles of dust and
synchrotron separately. In any case, differing from anisotropy measurements,
it seems very difficult to  measure free-free emission.

At this point, we can state that a conservative estimate of SPOrt full-sky
sensitivity to CBR polarization, arising from a single-pixel analysis and
taking into account a 50\% efficiency factor, is around 0.5--0.7 $\mu $K, the
uncertainty depending on the disturbance of free-free emission. However,
considering  free-free as a systematic effect
in $\sigma _{\mathrm{cbr}}$ is too pessimistic. Going
beyond the single-pixel analysis, we should exploit the spatial distribution
of the signal. Spatial information is often exploited also using
maps at other frequencies as
templates, or for some kind of spatial cross-correlations.
Basically we can
work in real space or in terms of modes
(for instance, using spherical harmonic expansions).

An important property of foregrounds is that they are more spatially
correlated than CBR. This is well known for the temperature anisotropy:
we have
$C_{Tl}\propto l^{-3}$
for free-free (from COBE--DMR: see (7,8)
) and dust (from IRAS (41))
, while $C_{Tl}\propto [l(l+1)]^{-1}$ for the standard
scale-invariant CBR spectrum. An important question however is, whether this
is true for polarized foregrounds, too. Sethi et al. (14) 
 found that
for dust 
\begin{eqnarray*}
T_0^2C_{El} &\simeq &8.9\times 10^{-4}l^{-1.3}\text{ (}\mu \text{K)}^2, \\
T_0^2C_{Bl} &\simeq &1.0\times 10^{-3}l^{-1.4}\text{ (}\mu \text{K)}^2,
\end{eqnarray*}
where the smaller (less negative) exponents mean smaller spatial
correlations than for anisotropies. However for CBR, too, the same
inequality applies; as a result, polarized foregrounds seem to be more
correlated than polarized CBR, as supported from the angular spectra
reported in Fig. \ref{angspct}. Provided this conclusion can be
generalized to the other foregrounds, a substantial improvement
should come from a spatial or spectral analysis.

It is therefore reasonable to believe that the contribution of $\sigma _{\mathrm shape}$
can be significantly reduced, and we expect to be able to set
$\sigma _{\mathrm cbr} \approx FDF \sigma ^{(0)}$.

\section*{Conclusions}

From the analysis of the previous section we can assume that SPOrt
effective sensitivity to CBR polarization will be around 0.4  $\mu $K, (cfr. the 3-rd line
in Table \ref{fdfsig}), or 0.5 $\mu $K including also the
Galactic cut. This
will allow us to probe a significant portion of parameter space for secondary
ionization models. In order to check
the validity of this conclusion, we performed extensive
computations of secondary ionization models using routines of the SPOrtLIB
library which is currently being built by the SPOrt collaboration.
\begin{figure}
\centerline{%
\epsfysize=7cm
\epsfbox{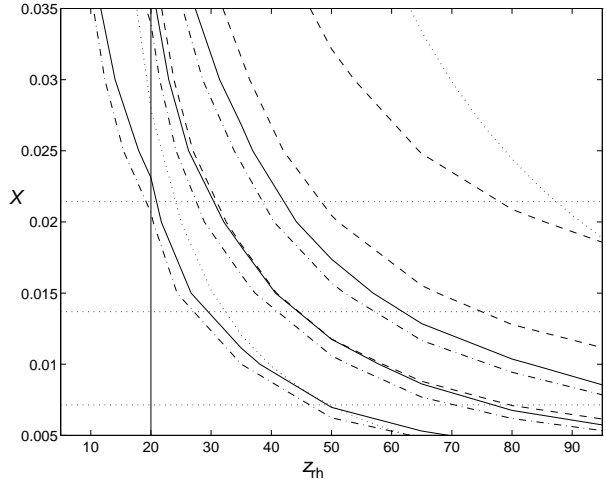}}
\caption{Polarization contours in the $(X, z_{\mathrm {rh}})$ plane corresponding 
to $7^\circ$ squared fluctuations of 0.1, 0.25 and 0.5 $\mu$K$^2$. The contours refer to CDM models
 with $\Omega_0=0.9$ and 0.3 (full and dashed lines, respectively) and a CDM+texture model 
with $\Omega_0=0.7$ (dash-dotted). The dotted curves give the constant optical depth contours 
for $\Omega_0=0.9$ and $\tau_{\mathrm rh}= 0.7$ and 0.1. The horizontal dotted lines give the baryon 
density limits in ref. (34) and the best value in ref. (37). The vertical line gives the best fit result of de Bernardis 
et al. (33).}
\label{figcnt}
\end{figure}
Some results are reported in
Fig. \ref{figcnt}, which provides polarization level contours
in the ($z_{\mathrm{rh}},X)$ plane for experiments with a $7^{\circ }$
beamwidth. The contours represent predictions from some 
cosmic structure models (with a
primordial spectral index $n=1$ and normalized to COBE--DMR spectral
amplitude $Q_{\mathrm{rms-PS}}=18$ $\mu $K) computed for several values of $%
\Omega _0$. For each value of $\Omega _0$ (i.e., for each line style) the
middle contour corresponds to the assumed full-sky sensitivity of 0.5 $\mu $%
K and thereby defines the region of the ($z_{\mathrm{rh}},X)$ plane where,
according to the assumed model of cosmic structure, the cosmological
polarization is accessible to SPOrt; such a region extends towards the upper
right corner of the Figure and is larger for $\Omega _0\simeq 0.7$. The
Figure also gives limits coming from the baryon density (for the assumed
value $h=0.7$) and some curves of constant $\tau _{\mathrm{rh}}$. Comparing
the locations of the model contours with the $\tau _{\mathrm{rh}}$ curves it
is clear that CBR polarization can be detected very  easily for optical depths $\sim 0.7$%
, i.e. around the upper limit derived from de Bernardis et al. (33
); for $\tau _{\mathrm{rh}}\sim 0.1$ SPOrt is likely to detect polarization
around its sensitivity limit.

\section*{Acknowledgements}

This work, as well as the SPOrt project,
is supported by Agenzia Spaziale Italiana (ASI). The European Space Agency
has supported SPOrt's A--B bridging phase under EPI industrial contracts.
M.V.S. thanks the CentroVolta--LandauNetwork for financial support. I.A.S.
thanks the Astronomy Dept. of the Bologna University and the ITeSRE/CNR
for the support given to his participation to SPOrt activities.

\end{document}